\documentclass[fleqn,usenatbib]{mnras}

\usepackage{mathptmx}

\usepackage[T1]{fontenc}
\usepackage[utf8]{inputenc}
\usepackage{ae,aecompl}

\usepackage{graphicx}	
\usepackage{amsmath}	
\usepackage{amssymb}	

\newcommand{\msun}{$\; \rm M_\odot$}

\title[UDGs in galaxy clusters]{The Formation of  Ultra-Diffuse
  Galaxies in Clusters}

\author[Sales et al.]{
\parbox[t]{\textwidth}{
Laura V. Sales$^{1}$\href{https://orcid.org/0000-0002-3790-720X}{\includegraphics[scale=0.8]{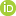}},
Julio F. Navarro$^{2}$,
Louis Pe\~nafiel$^{1,3}$\href{https://orcid.org/0000-0003-2494-2978}{\includegraphics[scale=0.8]{figs/orcid.png}} ,
Eric W. Peng$^{4,5}$\href{https://orcid.org/0000-0002-2073-2781}{\includegraphics[scale=0.8]{figs/orcid.png}}, 
Sungsoon Lim$^{6}$\href{https://orcid.org/0000-0002-5049-4390}{\includegraphics[scale=0.8]{figs/orcid.png}} 
and Lars Hernquist$^{7}$
}
\\
\\
$^{1}$University of California Riverside, 900 University Ave., Riverside CA 92521, USA\\
$^{2}$Department of Physics and Astronomy, University of Victoria, Victoria, BC V8P 5C2, Canada\\
$^{3}$Department of Physics, Cornell University, Ithaca, NY 14853, USA\\
$^{4}$Department of Astronomy, Peking University, 5 Yiheyuan Road, Beijing, China 100871\\
$^{5}$Kavli Institute for Astronomy and Astrophysics, Peking
University, 5 Yiheyuan Road, Beijing, China 100871\\
$^{6}$Herzberg Astronomy and Astrophysics Research Centre, National Research Council of Canada, Victoria, BC V9E 2E7, Canada\\
$^{7}$Harvard-Smithsonian Center for Astrophysics, 60 Garden Street, Cambridge, MA 02138, USA\\
}

\pubyear{2019}

\begin{document}
\label{firstpage}
\pagerange{\pageref{firstpage}--\pageref{lastpage}}
\maketitle

\begin{abstract}
  We use the IllustrisTNG cosmological hydrodynamical simulation to
  study the formation of ultra-diffuse galaxies (UDGs) in galaxy
  clusters. We supplement the simulations with a realistic mass-size
  relation for galaxies at the time of infall into the cluster, 
  as well as an analytical model to
  describe the tidally-induced evolution of their stellar mass,
  velocity dispersion and size. The model assumes ``cuspy'' NFW halos
  and, contrary to recent claims, has no difficulty reproducing the
  observed number of UDGs in clusters. Our results further suggest that the
  UDG population consists of a mixture of ``normal'' low surface
  brightness galaxies such as those found in the field (``born'' UDGs, or B-UDGs),
  as well as a distinct population that owe their large size and low
  surface brightness to the effects of cluster tides (``tidal'', or
  T-UDGs). The simulations indicate that T-UDGs entered the cluster
  earlier and should be more prevalent than B-UDGs near the cluster
  centres. T-UDGs should also have, at given stellar mass, lower
  velocity dispersion, higher metallicities, and lower dark matter
  content than B-UDGs. Our results suggest that systems like DF-44 are
  consistent with having been born as UDGs, while others such as DF2,
  DF4 and VLSB-D are possibly extreme T-UDG examples.
\end{abstract}




\section{Introduction}
\label{sec:intro}

Low surface brightness (LSB) galaxies have long been known to
constitute an important segment of the galaxy population, but
challenges in identifying them in large panoramic surveys have
resulted in incomplete counts and insufficiently accurate characterization of their
intrinsic properties \citep{Sandage1984,Dalcanton1997,Impey1988,Galaz2011,Greco2018}. In
the field, LSBs are most often systems of unusually large
size and relatively inefficient star formation that populate
predominantly the faint-end of the galaxy stellar mass function. Their
extreme properties make them as challenging to observe as they are to account
for in current models of galaxy formation.

The advent of LSB-sensitive surveys and telescopes with specialized
optics such as Dragonfly \citep{Abraham2014} have rekindled interest
in LSBs and in the faint tail of the surface brightness
distribution. New cluster surveys, together with tailored re-analysis
of existing photometric data, have resulted in the identification of a
``new'' population of dwarf (i.e., $M_*<10^9\, M_\odot$) LSBs in
clusters; these systems have spheroidal morphologies, are
predominantly gas free, and are mainly red in color. Using a
conventional definition of limiting effective surface brightness
$\mu_{\rm eff} \geq 25$ mag/asec$^2$ ($g$-band) and an effective radius
$R_e >0.7$-$1.5$ kpc these systems have
become known as ``Ultra-Diffuse Galaxies'' (UDGs)
\citep{vanDokkum2015,Koda2015,Yagi2016,Munoz2015,Mihos2015,
  vanderBurg2016, Venhola2017}.

Originally thought to be a distinct population, UDGs are now believed
to constitute the high-end of the size distribution of dwarf galaxies
in clusters and the field. Indeed, the $\mu_{\rm eff}>25$ mag/asec$^2$
criterion designates many ordinary LSB field galaxies as UDGs, a feature
that has often confused the discussion of whether UDGs are a ``cluster
phenomenon'' or just the faint end of the galaxy surface brightness
distribution.

Furthermore, systems of much fainter surface brightness
than any cluster UDG have long been known. For example, most
satellites of the Milky Way (MW) and Andromeda (M31) have much fainter
$\mu_{\rm eff}$ than any Dragonfly UDG in the Coma cluster.
Indeed, extreme
objects such as Crater II, And XIX, and Antlia II
\citep{Torrealba2016,McConnachie2008,Torrealba2019} have $\mu_{\rm eff}$ up to
$100\times$ fainter than Coma Dragonly
UDGs \citep[see; e.g.,][]{Fattahi2019}.

\begin{figure}
	\includegraphics[width=\columnwidth]{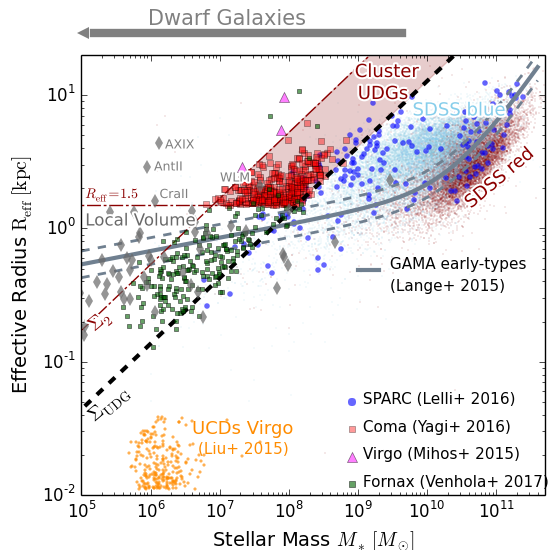}
        \caption{Effective radii $R_{\rm eff}$ as a function of
          stellar mass $M_*$ for a compilation of observed galaxies from
          the literature. Cyan and red points are used to indicate
          SDSS galaxies bluer and redder than
          $(g$-$r)_{\rm lim} = 0.61 + 0.052 (\log_{10}(M_*/M_\odot) -
          10.0)$, respectively. Galaxies from the SPARC compilation \citep{Lelli2016} are shown with large
          blue circles. The loci of GAMA early-type galaxies are shown by the grey
          curves, including scatter. Green and red squares indicate
          galaxies in the Fornax and Coma clusters, respectively
          (references in the legend). Yellow points indicate
          ultra-compact dwarfs in Virgo. UDGs are defined as in
          \citet{Koda2015}: these are galaxies with effective surface
          brightness ($\Sigma_{\rm eff}=(1/2)M_*/R_{\rm eff}^2$)
          lower than  $\Sigma_{UDG} \sim 2.9 \times 10^7 \, M_\odot$/kpc$^2$
          and higher than $\Sigma_2 \sim 0.2 \times 10^7 \, M_\odot$/kpc$^2$ 
	  (or, equivalently,  $\mu \sim 25$-$28$ $r$-mag/arcsec$^2$).  
          An additional
          criterion $R_{\rm eff}>1.5$ kpc is also introduced in order
          to exclude ``normal'' more compact dwarfs \citep{vanDokkum2015}. UDGs include data
          from \citet{Yagi2016,Venhola2017,Liu2015,Mihos2015}. 
	  Gray diamond-like symbols indicate Local Volume
          dwarfs, with highlighted systems of lower surface
          brightness than typical UDGs, such as  Antlia II, Crater II,
          Andromeda XIX and WLM.}
    \label{fig:obs}
\end{figure}

Although UDGs are clearly not just a cluster phenomenon, this does not
mean that clusters are not special environments that may radically
transform galaxies and somehow favor or induce the formation of
UDGs. Indeed, it is well established that dwarf galaxies in clusters
span a dramatic range in sizes (Fig.~\ref{fig:obs}), from the
ultra-compact dwarfs common in Virgo \citep[UCDs,][]{Liu2015} to the
UDGs in Coma \citep{Yagi2016}, Virgo \citep{Mihos2015}, and Fornax
\citep{Venhola2017}\footnote{We have assumed a stellar mass to light
  ratio $\gamma=1.96$ for $r$-band \citep{Carleton2019} and $\gamma=1$
  for $V$-band to estimate stellar masses.}. Explaining this enormous
range in dwarf galaxy sizes remains a difficult challenge for galaxy
formation models.

Regarding UDGs, three questions are still hotly debated: the mass of
their surrounding halos; the origin of their large sizes; and their
relation to the environment. Halo mass estimates are necessarily
indirect, but it has been argued that the high stellar velocity
dispersions of systems like DF44 and DFX1 indicate halos as massive as
$\sim 10^{12}\, M_\odot$, suggestive of ``failed'' Milky-Way like
galaxies \citep{vanDokkum2017}. 
On the other hand, several other UDGs
have lower velocity dispersions, consistent with the low mass halos
expected given their stellar mass \citep{Zaritsky2017, Toloba2018}.
More recently, improved data and modelling of some of the 
highest velocity dispersion UDGs have also favored lower mass halos of 
$\sim 10^{11}\, M_\odot$ \citep[e.g.][]{vanDokkum2019c}.

The globular cluster populations of UDGs are also quite diverse, and have
been used to argue for a wide range of halo masses
\citep{Beasley2016, Peng2016, Lim2018,Toloba2018}.  Finally, at least
2 UDGs have velocity dispersions so small that they may be consistent
with having no dark matter halo at all
\citep{vanDokkum2018a,vanDokkum2019}. This issue, and their
interpretation, has generated a healthy discussion that is still
ongoing \citep[see e.g., ][]{Martin2018,Laporte2019,Trujillo2019}.
Given these mixed properties, there is growing consensus that 
more than one formation path may be needed to explain the diversity of
UDGs.

Several UDG formation scenarios have been proposed, and may be grouped
as ``internal'' or ``external''. Internal scenarios include those
where UDGs correspond to the high-spin tail of the galaxy angular
momentum distribution \citep{Amorisco2016}, although this is
disfavoured by the spheroidal (i.e., non-disk) morphology of some
cluster UDGs \citep{Roman2017b,Eigenthaler2018} as well as by results of 
some hydrodynamical simulations \citep{Jiang2019,Tremmel2019}.
Another scenario posits that most dwarfs
go through expansion and contraction phases driven by sudden outflows
of gas following episodes of active star formation \citep[see;
e.g.,][]{Navarro1996,Pontzen2012}. Dwarfs that are caught at the
expansion phase during cluster infall may be ``frozen'' in that stage,
leaving behind a dwarf of unusually low surface brightness
\citep{Chan2018,DiCintio2017}. This is a viable (albeit perhaps
contrived) scenario that awaits confirmation from detailed simulations
that include the evolution of such dwarfs in the cosmological setting
of cluster formation.

The most promising external scenario concerns the transformation
of galaxies as a consequence of the mean tidal field of the
cluster. Simulations of tidal stripping of galaxies embedded in extended dark matter
halos show that this process may lead  to remnants of much lower
surface brightness than their progenitors
\citep{Penarrubia2008,Errani2015,Tomozeiu2016}. This transformation
may be aided by the loss of the gas component due to ram pressure with
the cluster's gaseous component \citep{Arraki2014,Frings2017,Yun2019}. The
final outcome are puffed-up, red, quiescent objects that resemble UDGs 
\citep{Safarzadeh2017,Roman2017b,Jiang2019}. 

The formation of UDGs via tidal stripping has recently been
studied by \citet{Carleton2019}, who argue that turning a typical
spheroidal galaxy into a UDG can only be accomplished if the dark
matter halos that host field dwarfs have large constant density
``cores'', unlike the cusps expected in the cold dark matter (CDM)
scenario. This is because dwarfs in cored halos are more easily
disturbed and evolve more dramatically in surface brightness than
dwarfs in cuspy, NFW halos \citep{Navarro1996,Navarro1997}. The
\citet{Carleton2019} study, however, did not consider normal LSB
galaxies as potential UDG progenitors, an assumption that is not
clearly justified and that may compromise the general applicability of
their conclusion.

We address these issues here by supplementing the cosmological
simulations of galaxy cluster formation from the IllustrisTNG
simulation suite \citep{Pillepich2018b} with an analytical model of
the tidal disruption of galaxies in the cluster potential. Our model
adopts the results of \citet{Errani2015} and, although it is similar
in approach to that of \citet{Carleton2019}, it also differs from it
in a few important aspects. For example, we shall consider the whole
field galaxy population as potential UDG progenitors, and not just
compact early-type galaxies. In addition, we shall use the baryonic
version of the TNG simulations, and not just the dark-matter-only run
of Illustris-1. This is important, as the loss of baryons in dwarf
galaxies (which can only be followed in hydrodynamical runs) leads to
less tightly-bound systems more prone to tidal disruption
\citep{Brook2015,Sawala2015,Sawala2016}. In addition, the assembly of
the central cluster galaxy deepens the potential and may enhance tidal
stripping \citep{Donghia2010,Garrison-Kimmel2017}.  Our analysis is
similar to that of \citet{Fattahi2018} (who focused on the effects of
tidal stripping on satellites of MW-like systems), but scaled up in
mass to the ``satellites'' of massive clusters.

This paper is organized as follows. Sec.~\ref{sec:method} introduces
the TNG simulations and galaxy samples and describes the tidal
evolution model applied in this work. Sec.~\ref{sec:results} presents
our main results, including the formation of UDGs and their
present-day predicted properties. In Sec.~\ref{sec:concl} we summarize
and discuss the implications of our results.

\section{Methods}
\label{sec:method}

\subsection{The simulations}

We use the cosmological hydrodynamical simulation Illustris-The Next
Generation \citep[TNG for
short;][]{Pillepich2018b,Nelson2018,Springel2018,Marinacci2018,Naiman2018,Nelson2019}. The
TNG suite of simulations includes several box sizes of varying
resolution, as well as runs with and without baryons.

In this work we use the TNG100-1 run, which is the highest resolution
box with 107 Mpc on a side that includes the treatment of baryons. The
simulations follow the evolution of $2 \times 1820^3$ DM and gas
elements from a starting redshift $z\sim 127$ until today. TNG assumes
a $\Lambda$ cold dark matter cosmology ($\Lambda$CDM) with parameters
consistent with Planck XIII results \citep{Planck2016}:
$\Omega_{\rm M}=$ $\Omega_{\rm dm} + \Omega_{\rm b} = 0.3089$,
$\Omega_{\rm b}=0.0486$, $\Omega_\Lambda = 0.6911$, Hubble constant
$H_0=67.74$ km s$^{\-1}$ Mpc$^{-1}$, and a power spectrum with
primordial index $n_s=0.9667$ normalized to $\sigma_8=0.8159$.

Gravity and baryons are followed using the {\sc arepo} code
\citep{Springel2010}. The baryonic treatment is largely based on the
code used for its predecessor simulation suite, Illustris
\citep{Vogelsberger2013, Vogelsberger2014b,Vogelsberger2014a, Genel2014,Nelson2015}, with modifications
described in \citet{Pillepich2018a}.

In brief, TNG follows self-consistently the heating and cooling of the
gas down to $10^4$ K with an effective equation of state for gas above the star formation threshold
$n_{\rm SFT} = 0.2 \; \rm cm^{-3}$. Such gas can turn into stars with a local
efficiency of $1\%$, after which stellar evolution models calculate
the mass and metallicity of the stellar particles.

Stellar feedback from winds of massive stars and supernova explosions
is coupled kinematically to the gas, with an efficiency that varies
according to the metallicity of the stars. AGN feedback including the
fast- and slow- accretion rates of gas onto supermassive black holes is
modeled with a mixture of thermal and kinetic energy deposition modes
following \citet{Weinberger2017}. Finally, ideal MHD is coupled to the
hydrodynamics to track the properties of magnetic fields
\citep{Pakmor2011, Pakmor2013}.

At the resolution level of TNG100-1, the particle masses are
$m_{\rm gas} \sim 1.4 \times 10^6$ and $m_{\rm dm}=7.5 \times 10^6$\msun\
for baryons and dark matter, respectively. The gravitational softening
for collisionless elements like dark matter and stars is
$\epsilon \sim 740$ pc or better. The highest resolution gas
elements can have gravitational softenings as small as $\sim 185$
(comoving) pc.

Galaxies in the simulations are identified using {\sc subfind}
\citep{Springel2001,Dolag2009} on Friends-of-Friends (FoF) groups
\citep{Davis1985} identified with a linking length set to $0.2$ the
mean interparticle separation. The time evolution of galaxies and
halos is followed by using the {\sc SubLink} merger trees
\citep{rodgomez2015}.

\subsection{Simulated galaxy clusters}
\label{ssec:sample}

We select the most massive halos in the TNG100-1 simulated volume,
with virial\footnote{We adopt a virial definition where the averaged
  spherical density within the virial radius, $r_{200}$, is $200$
  times the critical density for closure.} mass
$M_{\rm 200} \geq 10^{14}\, M_\odot$ at $z=0$. This criterion selects 14 objects
(the most massive has $M_{200}=3.79 \times 10^{14}$\msun) with masses
comparable to that estimated for the Virgo cluster.

We consider for our analysis all galaxies associated with the FoF groups of these
clusters that have a minimum stellar mass
$5 \times 10^7$\msun (or, equivalently, at least 25 stellar
particles). We also impose a similar minimum dark matter mass to
ensure that spurious self-gravitating baryonic clumps that are occasionally
identified by {\sc subfind} are excluded from our sample.  A total of
$4,850$ galaxies spread over 14 clusters make up the full simulated
galaxy sample.

\begin{figure}
	\includegraphics[width=\columnwidth]{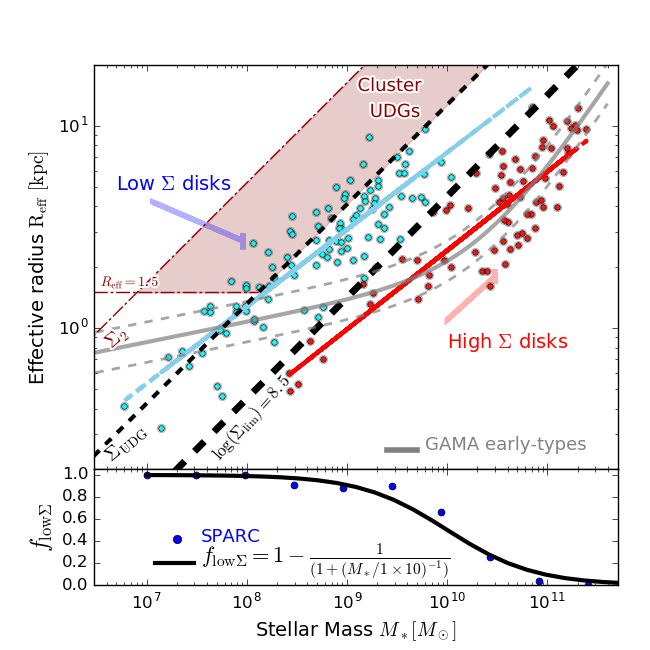}
        \caption{Field galaxies from the SPARC sample
          \citep{Lelli2016}. The sample may be split into a low (cyan)
          and high (red) surface density sequence, each following a
          slightly different $M_*$-$R_{\rm eff}$ relation, as fit by
          the cyan/red lines. The fraction of galaxies in each
          sequence as a function of $M_*$ is shown in the bottom
          panel. We use the stellar mass of infalling galaxies into
          simulated TNG clusters to assign them stellar sizes
          following these two sequences (adjusted by redshift of
          infall according to $(1+z_{\rm inf})^{-0.75}$, see text for
          details). For comparison, the early-type sequence from GAMA
          used by \citet{Carleton2019} to assign sizes is also
          indicated in gray.  The UDG selection criteria is
          the same as in Fig.~\ref{fig:obs}.}
    \label{fig:sparc}
\end{figure}

\subsection{The mass and size of simulated galaxies}

To study UDGs in the simulated galaxy population we need robust
estimates of stellar mass and half-mass radius. The TNG simulations
reproduce the galaxy stellar mass function remarkably well
\citep{Pillepich2018b}, suggesting that stellar masses are captured
correctly by the simulations. Galaxy sizes, on the other hand, are
much less robust, and are strongly affected by a number of numerical
artifacts and limitations, such as the use of a finite gravitational
softening, a limited number of particles, a relatively high threshold
for star formation, and an effective equation of state. Additionally,
the different particle mass used to model the dark matter and stars, 
a common practice in cosmological simulations, has been shown to cause 
an artificial transfer of energy to the stellar component via 2-body 
scattering that may also impact the predicted galaxy sizes \citep{Ludlow2019}. 
The combination of these effects mentioned above impose 
an effective minima on the size of simulated galaxies that affect dwarfs
in particular and that severely limit our ability to study UDGs
directly using simulation data.

We circumvent this limitation by assigning to each of our galaxies, at
the time of first cluster infall ($t_{\rm inf}$), a stellar
mass-dependent effective radius drawn from a realistic distribution of
galaxy sizes. Infall times are defined as the last time a galaxy
became a satellite of a larger system. This can be the time galaxies
join their clusters or an earlier time if they infall as part of a
galaxy group. Note that only the stellar radii are chosen that way;
stellar and dark halo masses are measured directly from the simulation
at $t_{\rm inf}$.  The tidal evolution model we describe below
(Sec.~\ref{ssec:tidalmodel}) allows us to compute their final stellar mass, effective
radii and velocity dispersion at $z=0$.

Effective radii are drawn assuming that they follow the
(redshift-corrected) empirical relations for late-type galaxies (including
LSBs) taken from the SPARC compilation of local, $z \sim 0$, 
galaxies \citep{Lelli2016}. This
assumption is consistent with the idea that before joining a larger
system, each galaxy is a star-forming galaxy in the field.

The stellar mass vs size relation of SPARC galaxies is shown in
Fig.~\ref{fig:sparc}, assuming a mass to light ratio of $0.5$ in the
3.6$\mu$ band \citep{Lelli2016}. This relation shows considerable scatter in the
projected stellar effective radius, $R_{\rm eff}$, at fixed stellar
mass. Two groups or ``sequences'' of ``high'' and ``low'' surface
brightness may be easily identified on each side of a dividing value
of $\Sigma_{\rm lim}=(1/2)M_*/R_{\rm eff}^2= 2.9 \times 10^8\; 
M_\odot$ kpc$^2$, indicated by the thick black dashed line.

The high-$\Sigma$ group dominates the massive end; the low-$\Sigma$
group contains most of the dwarfs. The fraction of galaxies that
belong to each sequence varies smoothly as a function of $M_*$, as
shown in the bottom panel Fig.~\ref{fig:sparc}. The low- and
high-$\Sigma$ sequences follow slightly different relations, which we
approximate as power laws: $\log_{10}(R_{\rm
  eff}/$kpc$)= a \log_{10}(M_*/M_\odot) + b$, with
$(a_{\rm low},b_{\rm low})=(0.38,-2.94)$ and
$(a_{\rm high},b_{\rm high})=(0.39,-3.51)$ respectively. These fits
are shown in Fig.~\ref{fig:sparc} with cyan and red lines,
respectively. The rms scatter in radii for both of these relations is
similar, $\sigma_{\rm log(r)} \sim 0.15$.

The grey curves in Fig.~\ref{fig:sparc} indicate the location of
early-type galaxies from the GAMA survey, including scatter
\citep[][]{Lange2015, Liske2015}. These overlap well the regime of
high-$\Sigma$ disks, but have much smaller sizes than the low-$\Sigma$
sequence. These low surface brightness objects are the most likely 
to turn into UDGs by
tidal effects, and leaving them out, as in \citet{Carleton2019}, who
only considered GAMA spheroids, can have a strong effect on the
results.

Our modeling uses the two-sequence description of SPARC galaxies to
assign sizes to all galaxies at infall time.  In practice, we proceed
as follows. Using its stellar mass at $t_{\rm inf}$, each galaxy is
randomly assigned to the low- or high-$\Sigma$ sequences by following
the SPARC fractions shown in the bottom panel of
Fig.~\ref{fig:sparc}. Once this sorting is done, we use the individual
power-law relations for the low- or high-$\Sigma$ sequence (plus
scatter) to assign an infall value of $R_{\rm eff}$ to each
galaxy. The projected effective radius is then converted into a 3D
stellar half mass radius, denoted with $r_h$ hereafter, by assuming:
$r_h = (4/3) R_{\rm eff}$. Although in principle this conversion
factor might change from $\sim 0.78$ in the case of elliptical galaxies 
to $\sim 1.2$ for disks \citep[see e.g., ][]{Somerville2018}, we choose
to keep the modeling simple by assuming the same transformation for
all galaxies regardless of their morphological type.

Additionally, because the infall redshift, $z_{\rm inf}$, can be quite
early, and galaxy sizes are known to decrease with increasing redshift 
\citep[see e.g., ][]{vanderWel2014}, we
correct the half mass radius $R_{\rm eff}$ (derived from the $z=0$ SPARC sample) 
to evolve at infall time by $(1 + z_{\rm inf})^{-0.75}$ \citep{Paulino-Afonso2017}.  This
procedure yields a population of cluster galaxies characterized by
their infall times and stellar masses (from the simulation) as well as
their sizes (computed as described above). The evolution of the dark
halos surrounding these galaxies may be followed to $z=0$ in order to
estimate the effects of cluster tides on the stellar component, as
described below.

\subsection{Tidal evolution model}
\label{ssec:tidalmodel}

\begin{figure}
	\includegraphics[width=\columnwidth]{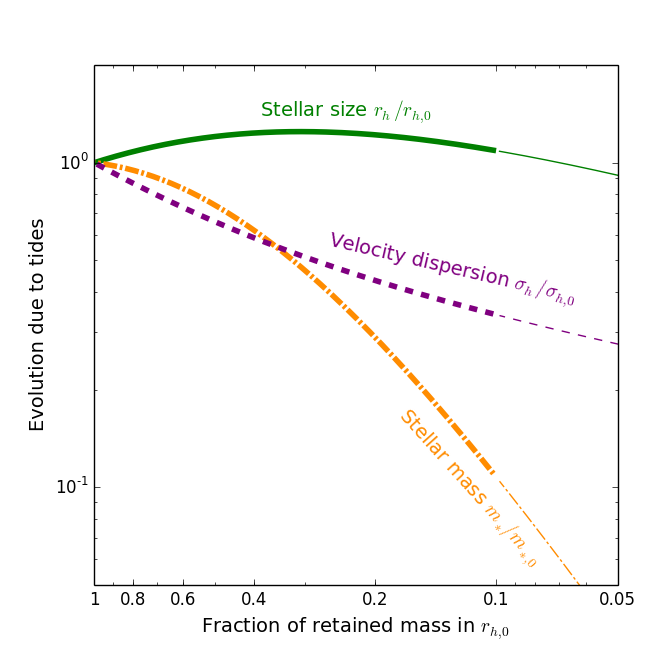}
        \caption{Evolution of stellar mass (orange), half-mass radius
          (green), and stellar velocity dispersion (purple) for a
          galaxy embedded in a cuspy NFW halo undergoing tidal
          disruption. Model taken from \citet{Errani2015}. }
    \label{fig:errani}
\end{figure}
Our tidal evolution model of cluster galaxies makes use of the work of
\citet{Penarrubia2008, Penarrubia2010} and \citet{Errani2015}, who showed that the
evolution of a stellar system deeply embedded in the gravitational
potential of a dark matter halo may be described in terms of a single
parameter. This parameter may be taken to be the {\it total} bound
mass fraction remaining within the original (3D) half mass radius of the
stars, $r_{h,0}$. Since each galaxy halo can be tracked through time, the total
mass lost by $z=0$ from within the initial (i.e., at infall) half-mass
radius can be computed directly from the
simulation. This mass loss can then be used to predict the changes in
stellar mass and effective radius of each individual cluster galaxy using the
effective radius assigned at infall. The mass loss estimates are robust, since
even the faintest galaxies in our sample inhabit fairly massive halos,
typically resolved with tens of thousands of dark matter
particles.

The tidal evolution model, which has been calibrated using numerical
simulations, may thus be used to predict the stellar half mass radius,
$r_h/r_{h,0}$, stellar mass, $M_*/M_{*,0}$ and velocity dispersion,
$\sigma_*/\sigma_{*,0}$, at the present day in units of their initial infall
values (``0'' subscript). The results are adequately approximated by  a simple formula:

\begin{align}
	g(x) = \frac{2^\alpha x^\beta}{(1 + x)^\alpha}
\label{eq:errani}	
\end{align}

\noindent
where $x$ is the fraction of mass that remains bound within $r_{h,0}$
and ($\alpha$,$\beta$) are constants given in Table~\ref{tab:errani}
and taken from \citet{Errani2015}. These constants vary slightly with the
relative segregation of stars relative to the dark matter. 
In what follows we assume the values with segregation $0.2$ in the \citet{Errani2015}
model, but we have checked that our conclusions are qualitatively unchanged if 
a segregation of 0.1 is assumed instead.

Fig.~\ref{fig:errani} illustrates the resulting tidal evolution tracks
when applying Eq.~\ref{eq:errani} as a function of $x$. Despite its
simplicity, these formulae have been shown to reproduce quite
well the tidal evolution of satellite galaxies in cosmological
hydrodynamical simulations of galaxy formation \citep[see, e.g, Fig.~3
of][]{Fattahi2018}. Additionally, using the dark matter
only run of TNG100 we have explicitly checked that the evolution of $V_{\rm max}$ and
$r_{\rm max}$ of satellites under tidal disruption in our clusters are well described 
by the tidal tracks from idealized simulations presented in \citet{Penarrubia2010}
within an r.m.s scatter $\sigma=0.04$ dex and $0.2$ dex for velocity and
$r_{\rm max}$, respectively.

Objects under tidal stripping may also suffer of ``tidal heating", by which
the stellar component may puff-up as a result of energy injection during 
the pericenter passages. This is included in the tidal tracks modelled by  
Eq.~\ref{eq:errani} and partially explain the slight expansion in radius
seen in Fig.~\ref{fig:errani}. However, as demonstrated by the comparison 
between the orange and green
curves in Fig.~\ref{fig:errani}, the main effect of tidal
stripping on the stellar component of a galaxy is to gradually reduce
its stellar mass while keeping its size approximately constant. This
implies that heavily stripped galaxies will be systems of much lower
surface brightness than expected at given stellar mass. Stripping also
reduces the stellar velocity dispersion, but to a somewhat lesser
extent.

The fits using Eq.~\ref{eq:errani} have only been validated by
simulation results for the range $0.1<x<1$. This range is indicated by
thick line types in Fig.~\ref{fig:errani}. Many of our cluster
galaxies have experienced mass losses that exceed the lower bound of
that range, with a sizable fraction ($5\%$) in the range
$0.01<x<0.1$. We shall assume that the same fitting formulae apply in
that regime, although we caution that the properties of such ``extreme
tidal remnants'' should be considered as preliminary until much higher
resolution simulations are available.

Finally, it is certainly possible that some cluster galaxies may have
experienced tidal losses even more extreme than $x\sim 0.01$, the
minimum we are able to track at the resolution of the TNG100-1
run. These systems would be missed in our simulated sample, which is
based on galaxies that survive\footnote{We note that recent work
  argues that current N-body simulations may substantially
  underestimate the number of surviving substructure in clusters
  \citep{vandenbosch2018a, vandenbosch2018b}.} to $z=0$, but they may
very well exist in true clusters. Their existence and, presumably,
their extreme properties cannot be discounted.

To summarize, we model the tidal evolution of individual galaxies by
measuring the total mass at $z=0$ within the half mass radius {\it
  assigned at infall} (as described in Sec.~\ref{ssec:sample}) and
comparing it with its value at infall. Eq.~\ref{eq:errani} then allows
us to predict the stellar mass, size and velocity dispersion at
present day of all cluster galaxies. We require a minimum of $10$ dark
matter particles at $z=0$ within the infall half-mass radius to
include a galaxy in our analysis.

\begin{table}
  \caption{Coefficients for the tidal evolution model in
    Eq.~\ref{eq:errani}, from \citet{Errani2015}.}
  \setlength{\tabcolsep}{5pt} 
  \begin {tabular*}{5.5cm}{{l} *{4}{c} }
    \hline
      &    &  $M_{*}/M_{*, 0}$  & $\sigma/ \sigma_0$  & $r_{\rm h}/
                                                        r_{\rm h, 0}$ \\
    \hline
    $\alpha$ &    & 3.57                        &  -0.68                &  1.22   \\
    $\beta$  &     & 2.06                        &  0.26                &  0.33    \\
    \hline
  \end{tabular*}
  \label{tab:errani}
\end{table}

\section{Results}
\label{sec:results}

\begin{figure*} 
\includegraphics[width=\columnwidth]{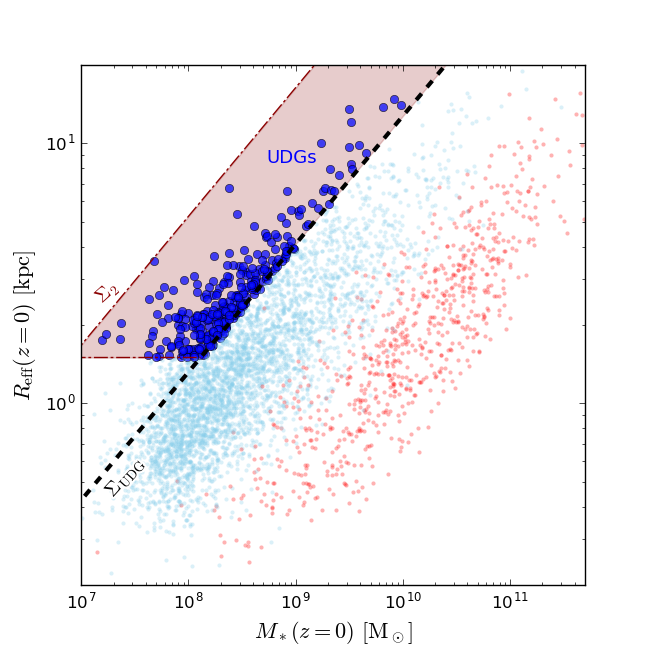}
\includegraphics[width=\columnwidth]{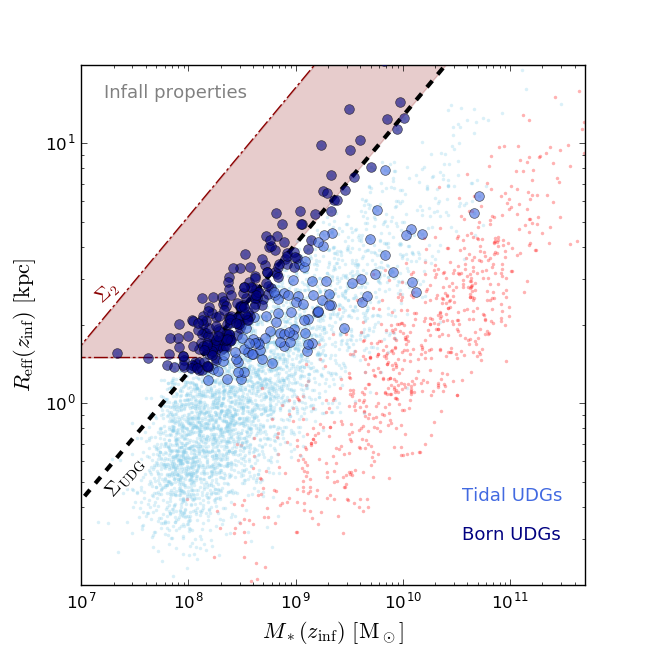}
\caption{{\it Left:} predicted stellar mass-size relation for
  simulated galaxies in$M_{\rm vir} \sim 10^{14}$\msun TNG clusters,
  evolved according to our tidal evolution model. Galaxies assigned
  originally to the low and high surface brightness sequences are
  shown in cyan and red, respectively. Systems consistent with a
  commonly used UDG definition are shown with large blue symbols
  within the shaded box. {\it Right}: same as left panel, but for the
  initial (i.e., at infall) $M_*$-$r_{\rm eff}$ relation of simulated
  galaxies. The population of UDGs at $z=0$ may be divided into two
  groups: galaxies that were originally 'born' consistent with the UDG
  definition (``Born UDGs'', dark blue) and galaxies that were more
  massive in the past and  evolve through tidal effects into objects with
  UDG properties today (``Tidal UDGs'', light blue).}
 \label{fig:mass_size}
 \end{figure*}
\subsection{UDGs in clusters} 

The result of the procedure described above, applied to all 14
clusters, is shown in the left panel of Fig.~\ref{fig:mass_size}. As
in Fig.~\ref{fig:sparc}, we use cyan and red symbols to indicate
galaxies that were assigned to the low- and high- surface brightness
sequences, respectively. The red shaded region indicate the UDG
criteria, defined as in \citet{Koda2015}: $\Sigma_{\rm eff}=(1/2)M_*/R_{\rm eff}^2 
< \Sigma_{\rm UDG}=2.9 \times 10^7 \, M_\odot$/kpc$^2$ (black dashed line), plus 
two additional boundaries: $\Sigma_{\rm eff}>\Sigma_2 > 0.2 \times
10^7 \, M_\odot$/kpc$^2$ (i.e., the equivalent
of r-band surface brightness 28.5 mag/arcsec$^2$) and $R_{\rm
  eff}>1.5$ kpc,
shown by dash-dotted red lines. The criterion
$R_{\rm eff}>1.5$ kpc is introduced in order to exclude
more compact, ``normal'' dwarfs \citep{vanDokkum2015}.  

On average,
$6$-$30$ galaxies fall in the UDG regime within the projected virial radius of
each cluster (a median of $17$ UDGs per cluster), and the numbers double if we consider the whole extent
of the cluster FoF groups, which extends slightly beyond the virial
boundaries of the cluster. Moreover, we have explicitly checked that
including scatter in the tidal tracks assumed for our analytical modeling
of tidal disruption would result in a somewhat larger number of UDGs per cluster
(for example, double the number of UDGs is formed when allowing for $0.2$ dex scatter
in the modeling of stellar size from Eq.~\ref{eq:errani}).

The main takeaway point is that UDGs have no trouble ``surviving'' in
the harsh tidal environment of a cluster in sizeable numbers. Indeed,
the number of simulated UDGs per cluster is in reasonable agreement
with the number of UDGs identified in Virgo (Lim et al., {\it
  in prep.}) (although these authors use a slightly different selection
criteria based on the overall shape of the $M_*$-size relation) 
as well as with the results of
\citet{vanderBurg2016,vanderBurg2017}, who report of order
$\sim 20$-$30$ UDGs for clusters of comparable virial mass.

We note that this result does {\it not} require that dark matter halos have
``cores'', as argued recently by \citet{Carleton2019}. Their analysis
and modeling is actually similar to ours, although they use the
dark-matter-only version of an earlier version of the Illustris
simulation suite. The main difference between our work and theirs is
that they only consider early-type galaxies from GAMA as potential
progenitors for UDGs. These galaxies are more compact than the SPARC
population we adopted here, and, hence, much more resilient to
stripping (see Fig.~\ref{fig:sparc} for a direct comparison). That
resilience is what led \citet{Carleton2019} to conclude that ``cored''
halos are needed to explain the UDG cluster population. We would argue
that our approach, which includes both low and high surface brightness
galaxies as potential UDG progenitors, samples more fairly  the field
population and is, therefore, more realistic.

\subsection{Origin of UDGs}

Where do the UDGs shown in the left-hand panel of
Fig.~\ref{fig:mass_size} originate from? This may be gleaned from the
cluster galaxy properties {\it at infall}, which are shown in the
right-hand panel of the same figure. Galaxies that end up as UDGs at
$z=0$ are shown with large blue circles. Two different hues are used;
dark to denote those that were within the UDG boundaries already at
infall, and a lighter hue for those that evolved to become UDGs as a
result of tidal stripping. In what follows, we refer to these two
populations as ``born UDGs'' (B-UDGs) and ``tidal UDGs'' (T-UDGs),
respectively. We emphasize that this classification 
corresponds only to the initial condition of each galaxy at infall 
and it does not mean that B-UDGs are free from tidal effects later
within the cluster. For T-UDGs, it is the stellar mass stripping at 
nearly constant radius (see Fig.~\ref{fig:errani}) 
that dominates their environmental transformation into UDGs.

The two panels of Fig.~\ref{fig:mass_size} illustrate a couple of
interesting points. One is that many cluster UDGs are relatively
unevolved remnants of field LSB galaxies that fell into the cluster
and, presumably, lost their gas and stopped forming stars. They may
have been affected by the cluster tidal field, but only mildly. This B-UDG
population make up $\sim 50\%$ of $z=0$ UDGs in our cluster simulation. A
second point is that the remainder, T-UDGs, make up a substantial
fraction of the UDG population, and originate from a varied mixture of
galaxies which were typically much more massive at infall than today. This mixed
origin can help to explain the wide spectrum of observed UDG properties
referred to in Sec.~\ref{sec:intro}. 

This is in good agreement with previous simulations
where approximately half of UDG-like 
objects form as simple analogs of dwarfs in the field (B-UDGs here) 
and the remaining half by tidal interactions in the form of tidal stripping 
or tidal heating \citep[][]{Jiang2019,Liao2019}. 
Our study extends these results in two main aspects. First, 
our field population follows, by construction, the observed 
stellar mass-size relation of galaxies and does not therefore 
overproduce the number of field UDGs. This is not the case for
\citet{Jiang2019}, where all simulated dwarfs with 
$M_* \sim 10^8\; \rm M_\odot$ are as diffuse as observed UDGs, 
at odds with the non-dominant fraction of dwarfs that are UDGs.
Second, we study UDGs in cluster environments
with $M_{\rm vir} \geq 10^{14}\; \rm M_\odot$, while previous studies
focused on groups \citep[$M_{\rm vir} \sim 10^{13.3} \; \rm M_\odot$, ][]{Jiang2019}
or Milky Way like hosts \citep{Liao2019}. It is, however, reasuring that
different techniques and simulation codes confirm the existence of more than one mechanism
to form UDG-like objects.

\begin{figure} 
\includegraphics[width=\columnwidth]{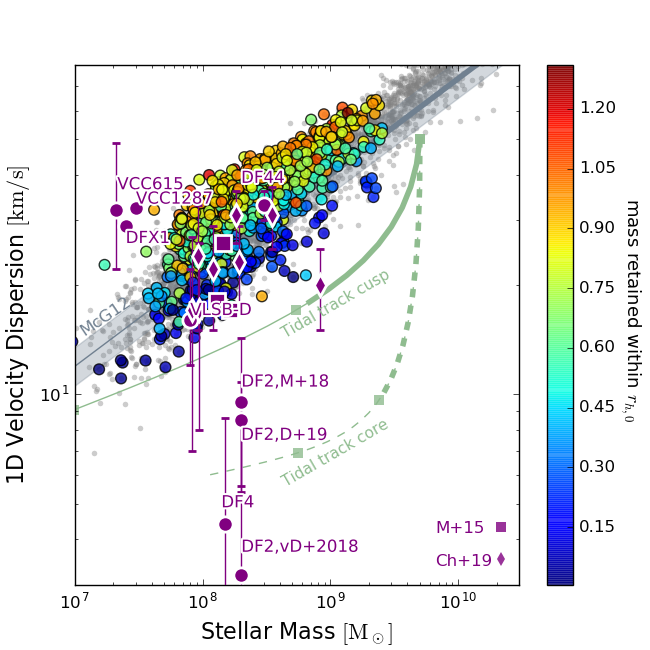}
\caption{Velocity dispersion-stellar mass relation
  for simulated galaxies in TNG clusters after accounting for their
  tidal evolution. For reference, the observed (baryonic) Tully Fisher
relation from \citet{McGaugh2012} is shown with a grey band, with rotation velocities scaled by
$\sqrt{3}$ to make them comparable to velocity dispersions. Candidate UDGs are color coded according to
  their retained bound mass fraction since infall (see color
  bar). UDGs that have retained most of their mass 
 (``born UDGs'') trace the general cluster galaxy population (shown in grey).
 ``Tidal UDGs'', i.e.,
  those formed by significant stripping of a more massive galaxy (low
  mass bound fractions---in blue), have systematically lower velocity
  dispersion at  given stellar mass. Data for a number of individual
  UDGs from the literature are
  shown in purple. While some UDGs such as Dragonfly-44
  are consistent with the ``born UDG'' scenario, others such as VLSB-D
  are candidates to be tidally-formed. See text for a more detailed
  discussion of extreme objects such as DF2 and DF4.}
\label{fig:udgs_sigma}
\end{figure}

\subsection{UDG velocity dispersions}

One of the most controversial aspects of UDGs has been their dark
matter content. This is inferred from the line-of-sight velocity
dispersion of stellar tracers, measured through spectroscopy of
stellar absorption lines, or from the kinematics of the globular
cluster systems. Published studies report a wide range of velocity
dispersions for UDGs, from $\sigma_{los} \sim 4$ km/s in objects like DF2
and DF4 \citep[so low that it may imply little to no dark matter,
][]{vanDokkum2018a,vanDokkum2019,Danieli2019} to as high as $\sim 50$
km/s for the globular cluster population ($\sim 33$ km/s for the stellar
dispersion at $R_{\rm eff}$) of DF44, which suggests a fairly massive dark matter halo
\citep{vanDokkum2016, vanDokkum2019_DF44}. 

We explore the velocity dispersion of simulated UDGs in
Fig.~\ref{fig:udgs_sigma}. Velocity dispersions are estimated from the
circular velocity at the stellar-half mass radius, assuming the
relation for dispersion-supported objects from \citet{Wolf2010}; i.e.,
$\sigma_{\rm los} = V_{\rm circ} / \sqrt{3}$. $V_{\rm circ}$ is
calculated directly from the particle data in the simulation by
measuring the dark matter and stellar mass enclosed within the evolved
($z=0$) half-mass radius of each galaxy. An additional 10\% downward
scaling is applied to all TNG velocities in order to correct for a
small systematic offset between the simulated and observed Baryonic
Tully-Fisher (BTF) relation of {\it field} galaxies \citep{McGaugh2012}.

Fig.~\ref{fig:udgs_sigma} shows the stellar mass-velocity dispersion
relation for all cluster galaxies at $z=0$ (grey points), as well as
for the UDGs (large circles, colored by the bound mass fraction
retained by each object). Reddish symbols correspond to
B-UDG systems, whose dark matter content within the stellar half mass
radius has been relatively unaffected by tides. Blue circles, on the
other hand, denote mainly T-UDGs, where tides have led to large mass
losses and, consequently, large changes in stellar mass and
velocity dispersion.

\begin{figure} 
\includegraphics[width=\columnwidth]{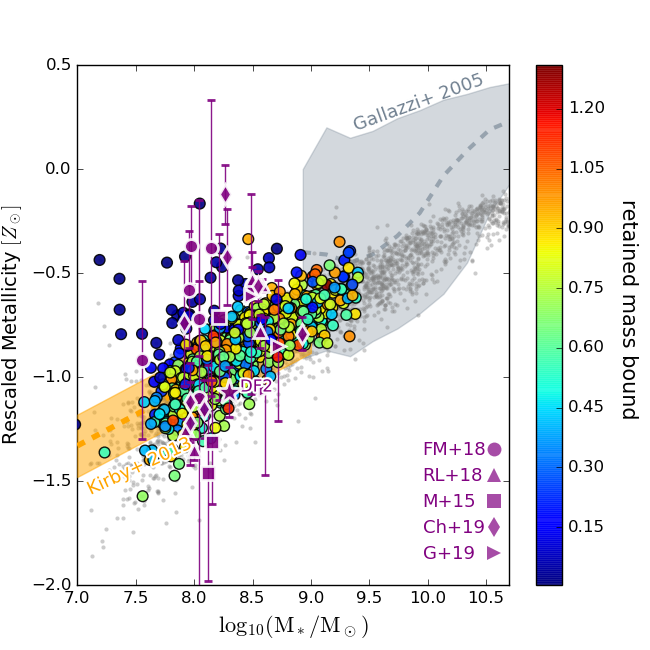}
\caption{Stellar metallicity-mass relation for simulated galaxies in
  TNG clusters after accounting for their tidal evolution. The color
  scheme is the same as in Fig.~\ref{fig:udgs_sigma}.  ``Tidal'' UDGs
  are outliers in this relation, with higher metallicities at given
  stellar mass than ``born UDGs" and the general cluster
  population. Our results indicate that the high metallicities
  measured for some UDGs \citep[see for instance
  ][]{Ferre2018,RuizLara2018} may be indicative of a tidal origin for
  these objects.}
\label{fig:udgs_metal}
\end{figure}

Cluster galaxies relatively unaffected by tides, as well as B-UDGs,
agree well with the observed BTF (grey shaded band labeled McG12 in
Fig.~\ref{fig:udgs_sigma}, where maximum rotation velocities have been
scaled by $\sqrt{3}$ factor to turn them into $\sigma_{\rm los}$). We
note that this agreement is not guaranteed by our procedure, which
only uses the stellar mass of a galaxy at infall to assign it a
stellar half-mass radius---its velocity dispersion follows mainly from
the total dark matter contained within that radius. Simulated galaxies
are slightly offset to higher velocities at fixed stellar mass or,
equivalently, to lower $M_*$ at fixed $\sigma_{\rm los}$.  This is
likely a consequence of our assumption that galaxies stop forming
stars at the time of infall, biasing galaxies to be less massive than
they should actually be.

Tidal losses can reduce both the stellar mass and the velocity
dispersion, pushing galaxies to the bottom of the relation seen in
Fig.~\ref{fig:udgs_sigma}. This provides one way of distinguishing
B-UDGs from T-UDGs: at fixed stellar mass, the latter should have
significantly lower velocity dispersions than the former.

Tidally-stripped galaxies move along ``tidal tracks'' in this figure,
such as the example  shown by the green solid curve in
Fig.~\ref{fig:udgs_sigma}. This curve show the loci of possible tidal remnants of a
galaxy progenitor with initial stellar mass, $M_{*,0}=5 \times 10^9 M_\odot$,
and 1D velocity dispersion $\sigma_{*,0}=50$ km/s.
Square symbols along the curve indicate remaining bound mass
fractions, spaced by successive factors of ten (i.e., $100\%$, $10\%$,
$1\%$, etc.).

For comparison, data for Coma and Virgo UDGs are shown with magenta
symbols and labeled individually in Fig.~\ref{fig:udgs_sigma} 
\citep{Toloba2018,vanDokkum2018a,vanDokkum2019,Martin2018,Danieli2019,vanDokkum2019_DF44,
Makarov2015,Chilingarian2019}.
High-velocity dispersion UDGs, such as DF44, are in agreement with
simulated B-UDGs. On the other hand, the comparatively low velocity
dispersion of the Virgo VLSB-D dwarf is in better agreement with
T-UDGs, which have undergone substantial tidal
disruption. Encouragingly, the morphology of VLSB-D is very elongated,
as is its globular cluster system, strongly suggestive of ongoing
tidal disruption \citep{Toloba2018}.

One main conclusion from this comparison is that the tidal origin of some UDGs implies substantial
diversity in velocity dispersion at fixed stellar mass. 
In addition, we note that the shape of the solid tidal track in
Fig.~\ref{fig:udgs_sigma} becomes shallower than the main
$M_*$-$\sigma_{\rm los}$ trend for extreme values of the mass
loss. This implies that 
velocity dispersions much below  $\sim 10$ km/s are highly unlikely
for systems with $M_*\sim 10^8\, M_\odot$. Such velocity dispersion is
at the high end of the estimates for DF2.  Much lower
velocity dispersions, such as those reported for DF4,
seem, at first glance, difficult to reproduce in our simulations,
where galaxies form in ``cuspy'', NFW-like halos
\citep{Bose2019}.  If confirmed, accommodating the lower limits
of the measured kinematics for DF2 and DF4 would be rather
challenging in our modeling. Indeed, few, if any tidal tracks originating in
the undisturbed galaxy population would leave a
remnant with such low velocity dispersion.

Several interpretations are possible for this result. One is that this
may indicate a non-tidal origin for these extreme objects; i.e., DF2
and DF4 may simply form in ways and/or halos with properties not
reproduced in our simulations. A more conservative interpretation,
however, is that the failure to reproduce systems like DF2 and DF4 is
a result of numerical limitations in the simulations, which prevent us
from including the most extremely tidally stripped systems in our
sample \citep[see ][for a similar conclusion]{Carleton2019}. Indeed,
we caution that the tidal tracks in Fig.~\ref{fig:udgs_sigma} include
a substantial extrapolation in the $x<0.1$ range, outside the regime
validated by available simulations (middle squared symbol in each
track; the extrapolated regime is shown as a thinner curve). It is
possible that larger drops in velocity dispersion may accompany
extreme tidal losses, but this t would need to be confirmed with
higher resolution simulations that extend the range probed in our
current tidal evolution model.

Finally, the extreme velocity dispersions of DF2 and DF4 may indicate
the presence of a ``core'' in the initial inner density profile of
their surrounding halos. Tidal tracks in the presence of a core differ
in shape, as indicated by the dashed green curve in
Fig.~\ref{fig:udgs_sigma}, which uses the results of the ``cored''
halo progenitor in the simulations of \citet{Errani2015}.  A core
leads to much lower velocity dispersions for extreme tidal remnants,
allowing better agreement with DF2 and DF4.

It is perhaps too early to use these results as evidence for a cusp or
core in such galaxies, especially given the limited tidal track range
actually validated by simulations and the large uncertainty in the
observational estimates, but it is an intriguing issue to which we
hope to return to in future work. Low velocity dispersion UDGs such as
DF2 and DF4 are akin to the ``cold faint giant'' satellites of the MW
and M31, Crater 2 and And XIX. The kinematics, size, and stellar mass
of the latter also suggest, in the context of current $\Lambda$CDM
models of dwarf galaxy formation, that they might be the remnants of
extreme tidal stripping events \citep{Fattahi2018}. Regardless of
whether halos have cores or cusps, these extreme tidal remnants should
be relatively rare. Should many systems comparable to DF2 and DF4 be
found, or if they were found in isolation (where tidal effects would
have been negligible), this would certainly call into question this
scenario.

\begin{figure} 
\includegraphics[width=\columnwidth]{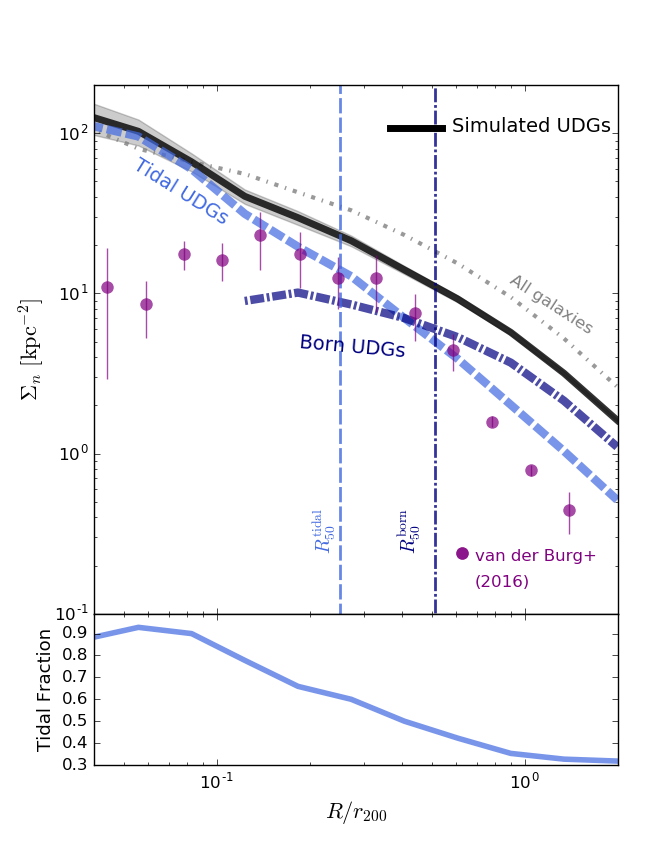}
\caption{Projected number density profiles for B-UDGs and T-UDGs.
  Tidally formed UDGs populate mainly in the central regions of
  clusters, whereas born UDGs have a more extended spatial distribution. The overall radial distribution of UDGs is not
  too different from that of all cluster galaxies, shown in grey (scaled by
  $\times 1/10$). For reference, observations for UDGs in Coma are shown using magenta symbols, 
  scaled down by a factor of $10$ to account for the difference in virial mass.}
  \label{fig:radprof}
\end{figure}

\subsection{UDG metallicities}

Like velocity dispersions, the metallicities of UDGs may also be used
to investigate their origin. In particular, T-UDGs are expected to deviate from
the mass-metallicity relation of B-UDGs, which should track the main
galaxy population. This is shown in Fig.~\ref{fig:udgs_metal}, where
it is clear that galaxies heavily affected by tidal striping (i.e.,
T-UDGs, those in bluish colors) may have, at fixed stellar
mass, higher metallicities than normal galaxies, or B-UDGs.

We note that simulations like TNG  generally have difficulty
matching the average mass-metallicity relation of dwarfs \citep[see ][for a more detailed
discussion]{Nelson2018}, so the simulated metallicities  in
Fig.~\ref{fig:udgs_metal} have been uniformly re-scaled to roughly match
$[Z]=-0.6$ at $M_*=10^9\, M_\odot$. This has little impact on our
conclusions, which refer mainly to the differential effect on the
metallicity of T-UDGs compared to normal galaxies(and B-UDGs)
similar stellar mass. Here, we have used the metallicity at $t_{\rm inf}$
for the simulated galaxies, but the same conclusion applies if using
the metallicities at $z=0$. For comparison, we include the present-day 
mass metallicity
relations from \citet{Gallazzi2005} and \citet{Kirby2013}, assuming a
solar abundance equivalent of $Z_\odot=0.0127$, as in \citet{Schaye2015}.

Available metallicity estimates for observed UDGs are included in 
Fig.~\ref{fig:udgs_metal} \citep{RuizLara2018,Ferre2018,Makarov2015,
Gu2018, Chilingarian2019,Fensch2019}
using magenta symbols, where we have transformed iron abundances 
[Fe/H] into full metallicities [Z/H] when necessary following Eq. 
3 in \citet{Boecker2019} and assuming [$\alpha$/Fe] = 0.25, 
the median [Mg/Fe] in the \citet{Ferre2018} sample.

Combining velocity dispersion with metallicity estimates provides an
excellent diagnostic of the effect of tides. Cluster (or satellite)
galaxies that have lower $\sigma_{\rm los}$ and higher metallicity
than expected for their stellar mass are excellent candidates for
being tidal remnants, or T-UDGs. The relatively low metallicity measured for 
DF2 \citep{Fensch2019} may complicate its interpretation as a tidal remnant,
although several simulated galaxies that have lost at least 55\% of their
mass within $r_{h,0}$ lie close to the  mass and metallicity of DF2. 
On the other hand, the relatively 
high metallicities of some of the measured UDGs, Yagi275 and Yagi276 
for example, bear well for our interpretation of such systems as extreme 
tidal remnants, although more detailed modeling is needed to validate 
this hypothesis.

\subsection{Radial distribution and orbital kinematics}

The spatial distribution and orbital kinematics of UDGs in clusters
also  carry  information about their origin, and it is something we
can readily study in our simulations. We explore this in
Fig.~\ref{fig:radprof}, where the upper panel shows the projected
number density profile of UDGs in all our 14 clusters. Clustercentric
radii are scaled to the virial radius, and three orthogonal
projections are used for each cluster before combining them to produce
the average density profile shown by the solid black line. For comparison, the
profile corresponding to all cluster galaxies (with $M_*>5 \times 10^{7} M_\odot$,
the minimum in our sample) is shown by the grey
dotted curve, which has been rescaled down by a factor of $10$ so as to
match the central density of the UDG curve. The comparison shows that
UDGs are only slightly biased relative to the cluster population at
large, with a somewhat steeper density profile.

Interestingly, B-UDGs and T-UDGs have quite different radial profiles:
the former tend to avoid the central regions of the cluster and are
overrepresented in the cluster outskirts, likely because they
typically inhabit low-mass halos that are fully disrupted
by tides in the central regions \citep[see also ][ for a similar
conclusion using analytical arguments]{Jiang2019}. There, only systems that were
originally fairly massive can survive the strong tidal field, leading
to a population of UDGs made up predominantly of T-UDGs. Half of all
T-UDGs are expected to lie inside one third of the virial radius,
whereas the half-number radius of B-UDGs is as large as
$\sim 0.5\, r_{200}$.

These results seem at odds with those reported by
\citet{vanderBurg2016} for the Coma cluster, where UDGs seem to avoid
the cluster center (see magenta symbols with error bars in
Fig.~\ref{fig:radprof}). Before reading too much into this
discrepancy, we note that UDGs are very difficult to identify
observationally, especially so near the cluster center, where the
intracluster light is especially bright and may compromise
detection. We plan to address this issue in more detail in future
work. We also note that UDGs in the Virgo cluster seem to follow a
cuspier distribution (Lim et al, in preparation).

\begin{figure} \includegraphics[width=\columnwidth]{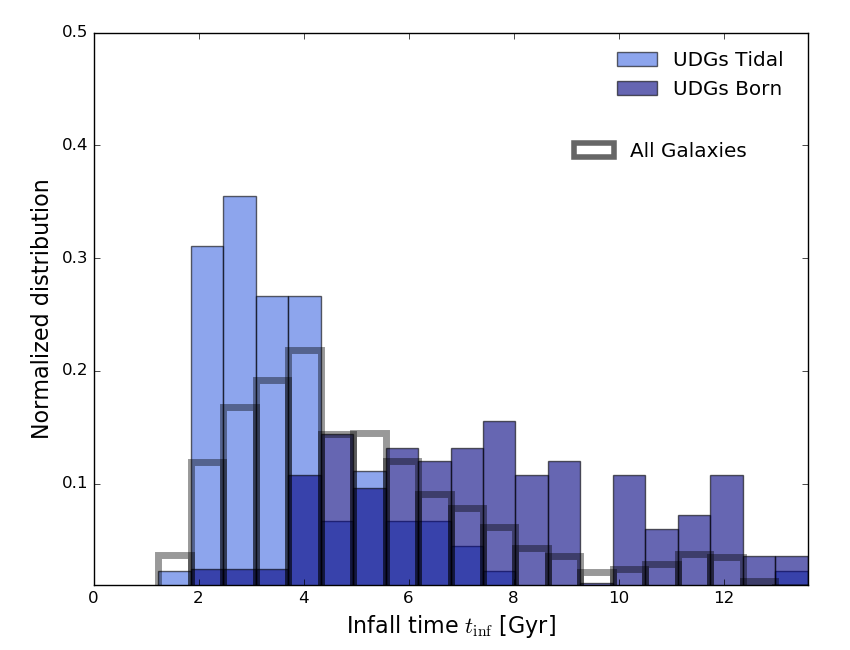}
  \caption{Cluster infall times of simulated galaxies. Tidal UDGs (light blue) are accreted preferentially at
    early times ($t_{\rm inf} < 6$ Gyr), consistent with their
    concentration near the cluster centres Born UDGs
    (navy blue) have later infall times compared with both
    Tidal UDGs and with the overall galaxy population (grey 
    histogram). }
  \label{fig:tinf}
\end{figure}

The distinct radial distribution of T-UDGs and B-UDGs is also
reflected in their infall times, which are shown in
Fig.~\ref{fig:tinf}. Both populations of UDGs, as well as the cluster
population as a whole, are shown with different color histograms,
normalized to the same area. B-UDGs infall much later than T-UDGs, as
expected. They also infall later than the population as a whole, which
indicates that most early infalling B-UDGs have been tidally disrupted and
are not present at $z=0$.

T-UDGs, on the other hand, are typically more massive systems at
infall (see Fig.~\ref{fig:mass_size}), are more resilient to tides,
and may survive to the present day. In some sense, this lends support
to the idea of UDGs as ``failed massive galaxies'', whose growth has
been truncated by cluster infall and whose properties have been
radically shaped by tides. The effect, however, is modest: the average
infall halo mass of $M_*=10^8\, M_\odot$ B-UDGs is roughly
$\sim 7 \times 10^{10}\, \rm M_\odot$, compared with
$2 \times 10^{11} \, \rm M_\odot$ for T-UDGs ($\sim 10\%$ of T-UDGs
have infall virial mass $\geq 5 \times 10^{11} \, \rm M_\odot$).

The infall time distinction between UDGs has interesting implications
for stellar ages. Assuming that most galaxies stop forming stars in
earnest soon after infall (as in our model), this implies that T-UDGs
should have much older stellar populations than B-UDGs. For instance,
\citet{Ferre2018} finds intermediate to old ages ($\sim 7$ Gyr)
stellar populations in Coma UDGs, and little difference with other
``normal'' dwarfs at similar cluster-centric distances. In our
scenario, this implies that they are mainly B-UDGs.

\begin{figure} \includegraphics[width=\columnwidth]{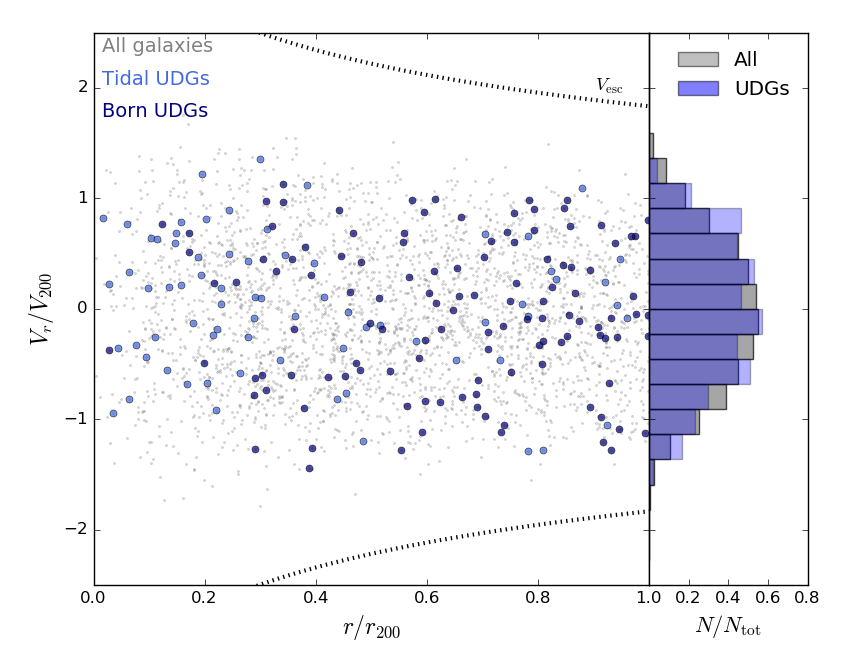}
  \caption{Radial velocity vs clustercentric distance $r$-$V_r$ 
    plot for simulated galaxies. The
    UDG population (Born- and Tidal-UDGs) has similar kinematics to
    the overall cluster population. Dotted
    curves indicate $V_{ \rm esc}$, the escape
    velocity of an NFW halo with mass and concentration consistent
    with the simulated clusters. For a more direct comparison, the vertical
    histograms show the velocity distributions for all galaxies (gray)
    and for simulated UDGs (blue) to be indistinguishable of each
    other. }
  \label{fig:r_vr}
\end{figure}

Since UDGs are well mixed with the general cluster population we expect
little kinematic differentiation between the two. This is shown in
Fig.~\ref{fig:r_vr}, where the phase diagram of all cluster galaxies
(scaled to virial quantities) is shown. Because the velocity dispersion is
nearly independent of radius, there is little difference in the
velocity distribution of T-UDGs, B-UDGs, and the cluster as a
whole. Most velocities sit comfortably within the escape velocity
boundaries of an NFW halo of concentration comparable to that of the
average cluster. Overall, these results seem to agree quite well with
those of \citet{Alabi2018}, who argue, based on their velocity
distribution, for a wide range of infall times for the UDG population.

\section{Summary and Conclusions}
\label{sec:concl}

We have used fourteen $\sim 10^{14}\, M_\odot$ galaxy clusters from
the Illustris TNG100-1 simulation to study the formation of cluster
UDGs in the $\Lambda$CDM cosmology. We supplement the simulations with
a tidal evolution model that allows us to track the structural
evolution of galaxies as they undergo tidal disruption in the
gravitational potential of the cluster. The model makes clear
predictions for the radial distribution and kinematics of UDGs, as
well as for the velocity dispersion, surface brightness, metallicity,
and ages of their stellar populations. We analyze the cluster
population at $z=0$ using similar UDG selection criteria as in
\citet{Koda2015}. We note that these criteria designate as UDGs a
number of field low surface brightness galaxies, which are likely
potential progenitors of cluster UDGs.

Our main finding is that cluster UDGs have primarily two different origins:
``normal'' field LSBs that are ``born'' as UDGs in the field and then
enter the cluster (B-UDGs), and a second population that is made up of
higher surface brightness galaxies transformed into UDGs by extensive
mass loss due to tides (T-UDGs). Tidal and born UDGs add up to a
sufficient number of UDGs to match the lower end of observational estimates for
$M_{200} \sim 10^{14}$\msun\ clusters \citep{vanderBurg2017}. We
conclude that there is no need to invoke ``cores'' in dark matter
halos to explain the abundance of cluster UDGs. 

Our analysis also identifies diagnostics that may be used to
distinguish between these two different formation paths. The
substantial mass loss needed to tidally form UDGs turns them into
outliers of typical galaxy scaling relations, with lower velocity
dispersions and higher metallicities at fixed stellar mass.  Tidal UDGs
also entered the cluster significantly earlier  ($\sim 9.5$ Gyr ago) than
surviving B-UDGs, which, on average, were accreted into the cluster
$\sim 5.5$ Gyr ago. As a result, tidal UDGs dominate in the inner regions
of galaxy clusters and their stellar ages should be on average older
than those of B-UDGs.

These results are, in principle, within reach of spectroscopic
studies. The few cases with robust metallicity/age estimates ($\sim
20$ objects to date) seem to have low metallicities ($[\rm
Fe/\rm H] \leq -1.1$), often enhanced alpha-elements ($[\rm
Mg/\rm Fe] \sim 0.4$), and intermediate to old stellar populations
$\sim 7$ Gyr \citep{Makarov2015, vanDokkum2016,
  Kadowaki2017,Ferre2018,RuizLara2018,Danieli2019}. These properties
are what would be expected from B-UDGs, which are in essence field
galaxies whose star formation history ceased after entering the
cluster but which were otherwise only modestly affected by cluster tides.

Our results lend some support to the ``evolutionary'' scenario
proposed by \citet{Roman2017b} that connects young UDGs in the field,
groups and clusters. However, our simulations also predict the
formation of an extra T-UDG population made up of galaxies severely
affected by tidal stripping. The large
metallicities of $2$ UDGs from the study \citet{Ferre2018}, Yagi275 and
Yagi276, suggest a possible tidal origin for these objects, a
result that deserves future investigation. 

Other examples of T-UDGs may be objects with low dark matter
component. A few candidates have been identified, such as VLSB-D
\citep{Toloba2018}, DF2 \citep{vanDokkum2018a} and DF4
\citep{vanDokkum2019}, with velocity dispersions \citep[as suggested
by their globular cluster system and stars for the case of DF2, see
][]{Danieli2019} that are too low to contain significant amounts of
dark matter.  Our model identifies these systems with ``extreme tidal
remnants'' that have lost the majority of their mass (dark and
luminous) to tides. If this is true, objects like VLSB-D, DF2, and DF4
should be rare, but they may yet provide the best evidence for the
substantial transformation of galaxies through tides that our models
predict.

The existence of a T-UDG population, with its predicted older ages,
enhanced concentration at the cluster center, lower velocity
dispersions, and higher metallicities is one of the key predictions
from our study, and one that future surveys, especially those able to
identify extreme LSB galaxies against the intracluster light near the
cluster center, should be able to validate or rule out.

\section*{Acknowledgments}
The authors would like to thank the anonymous referee for a 
constructive report that helped improve 
the first version of this manuscript. 
We would like to also thank Azadeh Fattahi, Christina Manzano-King 
and Cecilia Scannapieco for
making available some of the data included in Fig.~1. 
LVS acknowledges support from NASA through the HST Program AR-14583 and 
from the Hellman Foundation. EWP acknowledges support from the 
National Natural Science Foundation of China under Grant No. 11573002.
LH is grateful for support from NSF program AST1815978.




\bibliographystyle{mnras}
\bibliography{refs} 

\end{document}